# Synthesis, structure and properties of layered phosphide nitrides $Ak$Th$_2$Mn$_4$P$_4$N$_2$ ($Ak$ = Rb, Cs)

Bai-Zhuo Li,[a] Si-Qi Wu,[a] Zhi-Cheng Wang,[a] Cao Wang,[b] and Guang-Han Cao*,[a,c]

[a] *Department of Physics, Zhejiang Province Key Laboratory of Quantum Technology and Devices, Interdisciplinary Center for Quantum Information, and State Key Lab of Silicon Materials, Zhejiang University, Hangzhou 310027, China*
[b] *School of Physics & Optoelectronic Engineering, Shandong University of Technology, Zibo 255000, China*
[c] *Collaborative Innovation Centre of Advanced Microstructures, Nanjing University, Nanjing 210093, China*

**Main observation and conclusion**

We report the design, synthesis, structure, and properties of two complex layered phosphide nitrides, $Ak$Th$_2$Mn$_4$P$_4$N$_2$ ($Ak$ = Rb, Cs), which contain anti-fluorite-type [Mn$_2$P$_2$] bilayers separated by fluorite-type [Th$_2$N$_2$] layers as a result of the intergrowth between $Ak$Mn$_2$P$_2$ and ThMnPN. The new compounds are featured with an intrinsic hole doping associated with the interlayer charge transfer and a built-in chemical pressure from the [Th$_2$N$_2$] layers, both of which are reflected by the changes in the lattice and the atomic position of phosphorus. The measurements of magnetic susceptibility, electrical resistivity, and specific heat indicate existence of local moments as well as itinerant electrons in relation with $d$-$p$ hybridizations. The expected dominant antiferromagnetic interactions with enhanced $d$-$p$ hybridizations were demonstrated by the first-principles calculations only when additional Coulomb repulsions are included. The density of states at the Fermi level derived from the specific-heat analysis are 3.5 and 7.5 times of the calculated ones for $Ak$ = Rb and Cs, respectively, suggesting strong electron correlations in the title compounds.

**Comprehensive Graphic Content**

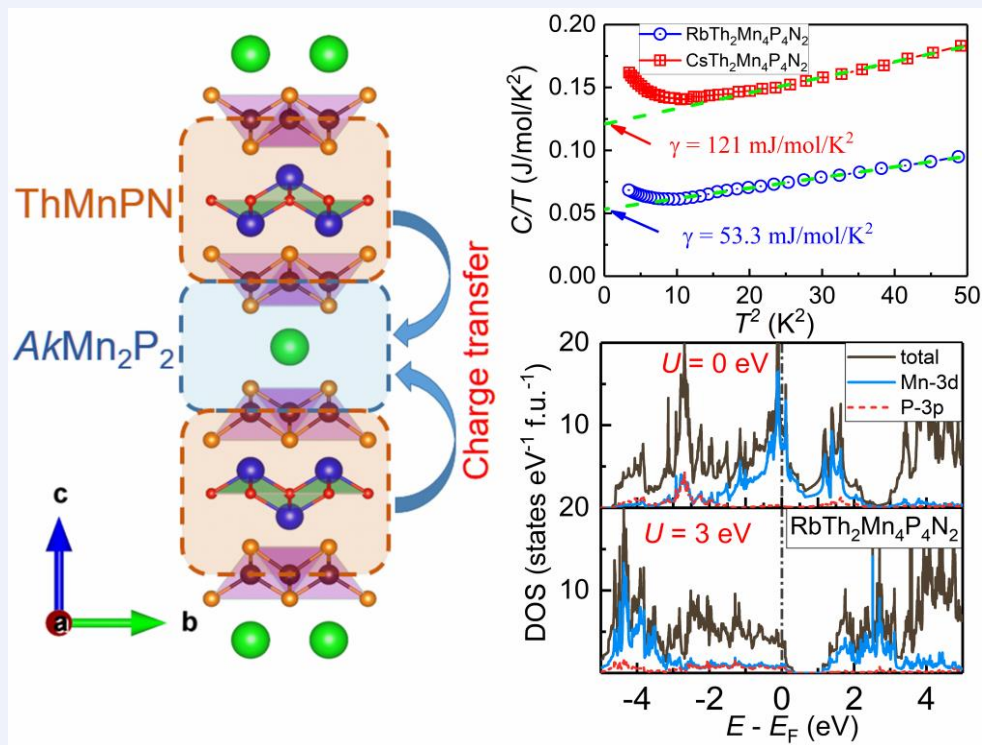

## Background and Originality Content

The discovery of high-temperature superconductivity in iron-based pnictides [1,2] stimulated investigations on the isostructural materials containing other 3$d$ transition metals [3-6]. Among them, Mn-based pnictides are unique due to the antiferromagnetic (AFM) insulating state in the parent compounds (with a formal valence state of $Mn^{2+}$) [4,7-14], akin to high-temperature superconducting cuprates. For example, LaMnPO is an AFM insulator with an optical band gap of ~1.3 eV and a Neel temperature of 375 K [4,8]. It was found that, unlike iron-based and cuprate systems, the AFM order in Mn-based pnictides is very robust against various chemical doping and high pressures [8,11-14]. Nevertheless, an insulator-to-metal transition was observed by a hole doping [12-14]. Besides, the metallization could also be achieved in LaMnPO by external pressures, yet the AFM order did not vanish unless very high pressures up to 30 GPa were applied [8,9]. Recently, we synthesized a new Mn-based compound ThMnPN, which bears an internal chemical pressure arising from the [$Th_2N_2$] layers [15]. As a result, this novel phosphide nitride appears to be intrinsically metallic with an AFM ordering at above 300 K. Notably, superconductivity at $T_c$ ≈ 1 K emerges in the binary compound MnP at a high pressure of 8 GPa [16]. It is thus of great interest to synthesize charge-carrier-doped MnP-based compounds with a built-in chemical pressure.

We note that charge-carrier doping can also be realized by an interlayer charge transfer in an intergrowth structure [17-22]. For example, the emergence of superconductivity in stoichiometric $KCa_2Fe_4As_4F_2$ is due to the intrinsic hole doping via an inter-block-layer charge transfer [17]. This so-called 12442-type material is formed by an alternate stacking of 1111-type CaFeAsF block and 122-type $KFe_2As_2$ block along the $c$ axis. The interlayer charge transfer not only generates charge carriers in the conducting layers, but also serves as the "glue" for the block-layer stacking, which makes the intergrowth structure thermodynamically stabilized [21,22]. These results inspire us to design and synthesize a similar intergrowth structure with both anti-fluorite-type [$Mn_2P_2$] layers and fluorite-type [$Th_2N_2$] layers.

Lattice match between distinct block layers also plays an important role in forming an intergrowth structure [20,22]. For a successful synthesis, empirically, the lattice mismatch, defined by $\mu = 2(a_1 - a_2) / (a_1 + a_2)$, where $a_1$ and $a_2$ are respectively the lattice parameter $a$ of the two constituent compounds, should be smaller than 2%. Now that the $a$ axes of $Ak$Mn$_2$P$_2$ are 4.0612 Å and 4.0948 Å for $Ak$ = Rb and Cs, respectively [23], and the $a$ axis of ThMnPN is 4.0301 Å [15], the criterion of $\mu < 2\%$ is thus satisfied. Additionally, the apparent valence of Mn in $Ak$Mn$_2$P$_2$ is 2.5+, while the valence state of Mn in ThMnPN is normally 2+. Therefore, an inter-block-layer charge transfer is expected, which could stabilize the target $Ak$Th$_2$Mn$_4$P$_4$N$_2$ designed.

In this paper, we report our successful synthesis of Mn-based phosphide nitrides $Ak$Th$_2$Mn$_4$P$_4$N$_2$ ($Ak$ = Rb, Cs), which are intrinsically hole-doped and bear an internal chemical pressure. The crystal structure shown in Fig. 1(c) is indeed an intergrowth between $Ak$Mn$_2$P$_2$ and ThMnPN, resulting in $Ak$-cation-connected conducting [$Mn_2P_2$] bilayers separated by the insulating [$Th_2N_2$] layers. Unusual changes in the lattice with the adjustments of atomic positions are found, which is primarily due to the inter-block-layer charge transfer. The comprehensive analyses of the data of electrical resistivity, magnetic susceptibility, and specific heat, together with the first-principles calculations, point to a novel local-moment AFM metallic state with significant electron correlations in $Ak$Th$_2$Mn$_4$P$_4$N$_2$.

## Results and Discussion

### Results

The target compounds $Ak$Th$_2$Mn$_4$P$_4$N$_2$ ($Ak$ = Rb, Cs) were synthesized by high-temperature solid-state reactions in evacuated containers. Details are presented in the following Experimental Section. The final products were characterized by powder XRD (Figs. 1(a) and (b)), which indicates successful syntheses of the expected 12442-type compounds. Small amount of the secondary phase Th$_3$P$_4$ can be detected, and the content is below 5% according to the relative intensity of the strongest reflections. The crystal structure was refined by the Rietveld analyses [24]. The occupation factors of all the constituent atoms were fixed to 1.0, according to the compositional analysis result with energy-dispersive X-ray spectroscopy (not shown here). The weighted $R$ factor and the "goodness of fit" parameter are $R_{wp}$ = 5.85% (6.35%) and $S$ = 1.21 (1.61), respectively, for RbTh$_2$Mn$_4$P$_4$N$_2$ (CsTh$_2$Mn$_4$P$_4$N$_2$), indicating good reliability of the crystal structure refined.

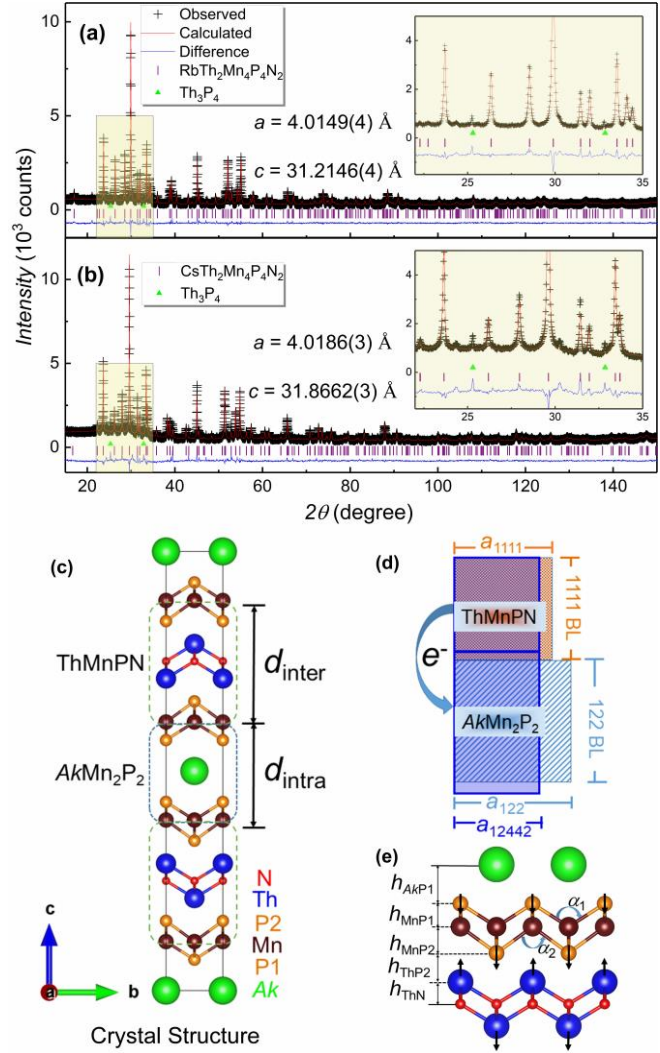

**Figure 1** Powder XRD and their Rietveld refinement profiles of the synthesized RbTh$_2$Mn$_4$P$_4$N$_2$ (a) and CsTh$_2$Mn$_4$P$_4$N$_2$ (b). Panel (c) shows the 12442-type unit cell projected along the [100] direction. Panel (d) sketches the lattice deformation due to the inter-block-layer charge transfer. Panel (e) illustrates the structural parameters listed in Table 2 in relation with the atomic position adjustments marked by the arrows.

The fitted crystallographic data are listed in Table 1. Surprisingly, the $a$ axes of RbTh$_2$Mn$_4$P$_4$N$_2$ (4.0149(4) Å) and CsTh$_2$Mn$_4$P$_4$N$_2$ (4.0186 Å) are both noticeably smaller than those of RbMn$_2$P$_2$ (4.0612 Å), CsMn$_2$P$_2$ (4.0948 Å), and ThMnPN (4.0301(3) Å). The result contrasts with the previous observations, which show the $a$ axis of the intergrowth phase always lies in between those of the constituent compounds [19,22]. Meanwhile, the $c$ axis of $Ak$Th$_2$Mn$_4$P$_4$N$_2$ are 1.1% (0.9%) larger than the expected values ($c_{122} + 2c_{1111}$) for $Ak$ = Rb and Cs, respectively. The lattice-size changes of the constituent structure blocks are sketched in Fig. 1 (d). One sees that the 1111 block shrinks both in the $ab$ plane and in the $c$ direction, while the 122 block is remarkably stretched along the $c$ axis. The pronounced size change of the 122 block is due to its smaller rigidness [22] and, because of the much higher rigidness of the ThMnPN block, the $a$ axis of $Ak$Th$_2$Mn$_4$P$_4$N$_2$ is dominantly determined by the size of 1111 block. The shrink in the 1111 block should be attributed to the inter-block-layer charge transfer, because the $a$ axis of the related compounds generally decreases with hole doping, which is explicitly shown in Ba$_{1-x}$K$_x$Fe$_2$As$_2$ system [25].

Table 2 compares some structural parameters of $Ak$Th$_2$Mn$_4$P$_4$N$_2$ and the related compounds, from which adjustments of the atomic positions in the $c$ direction can be traced. As shown in Fig. 1(e), the elongation of the 122 block comes from the remarkable increase of $h_{AkP1}$, and the compression of the 1111 block (decrease of $d_{inter}$) is caused by the decline of $h_{ThP2}$. Note that $d_{inter}/a$ values of RbTh$_2$Mn$_4$P$_4$N$_2$ and CsTh$_2$Mn$_4$P$_4$N$_2$ (2.151 and 2.152, respectively) are very close to the $c/a$ value in ThMnPN (2.155) [15], suggesting that the chemical pressure from the [Th$_2$N$_2$] layers holds. The change in $h_{ThP2}$ suggests that the transferred holes at least partially reside in the P2 atoms. We will come back to this issue later on.

**Table 1** Crystallographic data of $Ak$Th$_2$Mn$_4$P$_4$N$_2$ at 300 K ($Ak$ = Rb, Cs) obtained by the Rietveld refinement of the powder XRD data. The space group is $I4/mmm$ (No. 139).

|  |  | RbTh$_2$Mn$_4$P$_4$N$_2$ |  |  | CsTh$_2$Mn$_4$P$_4$N$_2$ |  |
|---|---|---|---|---|---|---|
| $a$ (Å) |  | 4.0149(4) |  |  | 4.0186(3) |  |
| $c$ (Å) |  | 31.2146(4) |  |  | 31.8662(3) |  |
| $V$ (Å$^3$) |  | 503.168(1) |  |  | 514.617(1) |  |
| $R_{wp}$ (%) |  | 5.85 |  |  | 6.35 |  |
| $R_p$ (%) |  | 4.46 |  |  | 4.55 |  |
| $S$ |  | 1.21 |  |  | 1.61 |  |
| *Atoms* | *Wyckoff* | *Occ.*(fixed) | *x* | *y* | *z* | $B_{iso}$ |
| | | RbTh$_2$Mn$_4$P$_4$N$_2$ | | | | |
| Rb | 2a | 1.0 | 0 | 0 | 0 | 1.8(5) |
| Th | 4e | 1.0 | 0.5 | 0.5 | 0.2106(1) | 0.2(1) |
| Mn | 8g | 1.0 | 0.5 | 0 | 0.1110(1) | 1.8(1) |
| P1 | 4e | 1.0 | 0.5 | 0.5 | 0.0698(3) | 0.8(2) |
| P2 | 4e | 1.0 | 0.5 | 0.5 | 0.1581(1) | 0.2(5) |
| N | 4d | 1.0 | 0.5 | 0 | 0.25 | 1.5(5) |
| | | CsTh$_2$Mn$_4$P$_4$N$_2$ | | | | |
| Cs | 2a | 1.0 | 0 | 0 | 0 | 1.2(4) |
| Th | 4e | 1.0 | 0.5 | 0.5 | 0.2116(1) | 0.3(1) |
| Mn | 8g | 1.0 | 0.5 | 0 | 0.1143(1) | 1.2(3) |
| P1 | 4e | 1.0 | 0.5 | 0.5 | 0.0732(4) | 0.2(4) |
| P2 | 4e | 1.0 | 0.5 | 0.5 | 0.1597(2) | 0.2(5) |
| N | 4d | 1.0 | 0.5 | 0 | 0.25 | 0.6(4) |

**Table 2** Structural parameters of RbTh$_2$Mn$_4$P$_4$N$_2$ (Rb-12442) and CsTh$_2$Mn$_4$P$_4$N$_2$ (Cs-12442) in comparison with ThMnPN and CsMn$_2$P$_2$ (Cs-122). $d_{MnP1}$ is the bond distance of Mn−P1, $h_{MnP1}$ refers to the P1 height over the Mn plane, $\alpha_1$ denotes the band angle of P1−Mn−P1, …, all of which are illustrated in Figs. 2(c) and (e).

|  | Cs-122 | ThMnPN | Rb-12442 | Cs-12442 |
|---|---|---|---|---|
| $d_{MnP1}$ (Å) | 2.385 | -- | 2.385(4) | 2.399(6) |
| $d_{MnP2}$ (Å) | -- | 2.453(2) | 2.488(5) | 2.478(6) |
| $d_{intra}$ (Å) | 7.102 | -- | 6.987(7) | 7.285(7) |
| $d_{inter}$ (Å) | -- | 8.684(7) | 8.637(7) | 8.650(7) |
| $h_{MnP1}$ (Å) | 1.591 | -- | 1.287(3) | 1.310(4) |
| $h_{MnP2}$ (Å) | -- | 1.394(2) | 1.469(3) | 1.449(4) |
| $\alpha_1$ (°) | 118.24 | -- | 114.7(3) | 113.8(5) |
| $\alpha_2$ (°) | -- | 110.49(1) | 107.6(3) | 108.4(4) |
| $h_{AkP1}$ (Å) | 1.960 | -- | 2.197(3) | 2.332(5) |
| $h_{ThP2}$ (Å) | -- | 1.764(2) | 1.621(3) | 1.651(4) |
| $h_{ThN}$ (Å) | -- | 1.179(2) | 1.228(2) | 1.223(3) |

Note that the [Mn$_2$P$_2$] layers in $Ak$Th$_2$Mn$_4$P$_4$N$_2$ are asymmetric (absence of local S4 symmetry), unlike $Ln$MnPO ($Ln$ = La, Ce, Pr, Nd, Sm, Eu, Gd, Tb, Dy) [26] or $A$Mn$_2$P$_2$ ($A$ = Rb, Cs, Ba) [7,23]. According to Table 2, the bond distances of Mn−P1 and Mn−P2 differ. As an example for CsTh$_2$Mn$_4$P$_4$N$_2$, the P1 height ($h_{MnP1}$) decreases to 1.310 Å (from 1.581 Å for CsMn$_2$P$_2$). Meanwhile, $h_{MnP2}$ increases to 1.449 Å (from 1.394 Å for ThMnPN). That is to say, P1 atoms move towards the Mn plane, and P2 atoms move away from the Mn plane. The result reflects the inter-block-layer charge transfer at the "interface" of the two building blocks.

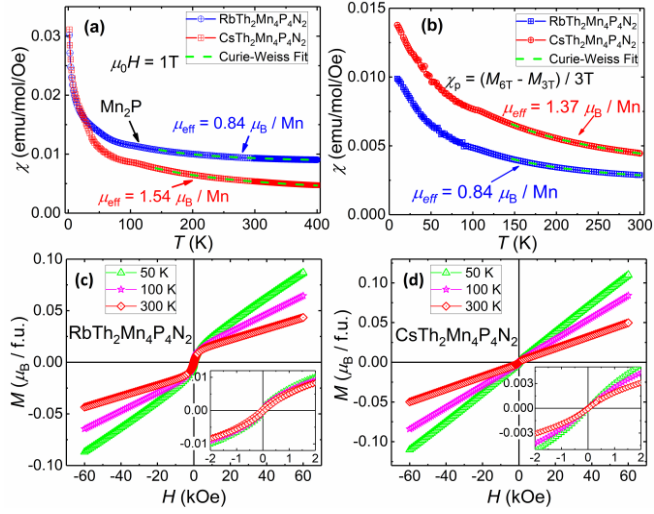

**Figure 2** Temperature dependence of magnetic susceptibility, defined by $\chi = M/H$ (a) and $\chi_P = (M_{6T} - M_{3T})/30000$ (b), for the $Ak$Th$_2$Mn$_4$P$_4$N$_2$ ($Ak$ = Rb, Cs) samples. The dashed lines are the Curie-Weiss fits, which give the effective magnetic moments shown. The bottom panels (c) and (d) show the field dependence of magnetization at fixed temperatures. The insets are the closeup for showing the weak ferromagnetism.

Figures 2(a) and (b) show the magnetic measurement results of the two samples of $Ak$Th$_2$Mn$_4$P$_4$N$_2$. The temperature dependence of magnetic susceptibility shows a Curie-Weiss (CW) paramagnetic behavior. The slight anomaly at 105 K is attributed to the AFM transition of tiny Mn$_2$P impurities (not detected by XRD) [27]. It is noted that the possible ferromagnetic MnP impurity (with a Curie temperature of 292 K [16]) does not exist in the samples because there is no rise of susceptibility at about 290 K. However, the $MH$ curve at 300 K shows ferromagnetic-like magnetization with very small saturation magnetization of ~0.01 and ~0.003 $\mu_B$/f.u. for $Ak$ = Rb and Cs respectively, as shown in Figs. 2 (c) and (d). The spontaneous magnetization is too small to be ascribed to an itinerant ferromagnetism [28,29]. Alternatively, the weak ferromagnetism could be due to an AFM transition with tiny spin canting due to the competition between the super- and double-exchange interactions [30]. The spin canting in RbTh$_2$Mn$_4$P$_4$N$_2$ is more pronounced, suggesting a stronger double-exchange interaction which is confirmed by the metallicity (see below).

The CW fit in the temperature range of 150 K < $T$ < 400 K, shown in Fig. 2(a), using the formula $\chi = \chi_0 + C/(T - \theta)$ yields $\chi_0 = 8.1 \times 10^{-3}$ emu mol$^{-1}$, $C = 0.355$ emu K mol$^{-1}$, and $\theta = 11.7$ K for RbTh$_2$Mn$_4$P$_4$N$_2$ and, $\chi_0 = 2.2 \times 10^{-3}$ emu mol$^{-1}$, $C = 1.19$ emu K mol$^{-1}$, and $\theta = -79$ K for CsTh$_2$Mn$_4$P$_4$N$_2$. Note that the $\chi_0$ value of RbTh$_2$Mn$_4$P$_4$N$_2$ is much larger than that of CsTh$_2$Mn$_4$P$_4$N$_2$ because of the pronounced ferromagnetic component. In fact, the intrinsic paramagnetic susceptibility $\chi_P$ should take the slope of the linear part of each $MH$ curve. We then extracted $\chi_P$ with the formula $(M_{6T} - M_{3T})/30000$, which is shown in Fig. 2(b). The CW fit gives $\chi_0 = 1.7 \times 10^{-3}$ and $1.75 \times 10^{-3}$ emu mol$^{-1}$ for $Ak$ = Rb and Cs, respectively. The positive $\chi_0$ values suggest contribution from Pauli paramagnetism in $Ak$Th$_2$Mn$_4$P$_4$N$_2$. With the fitted Curie constants, the effective local magnetic moment is calculated to be 0.84 and 1.37-1.54 $\mu_B$/Mn for $Ak$ = Rb and Cs, respectively, which are much reduced compared with the high-spin value of 5.92 $\mu_B$/Mn. Note that a similar reduced effective moment of 1.36 $\mu_B$/Mn is also observed in ThMnPN where there is an AFM transition at above 300 K. In fact, most Mn-based pnictides show AFM transitions at above room temperature. Examples include LaMnPO ($T_N$ = 375 K) [4,8], BaMn$_2$As$_2$ ($T_N$ = 625 K [10] or 613 K [31]), BaMn$_2$P$_2$ ($T_N$ > 750 K) [7], and NdMnAsO ($T_N$ = 359 K) [32]. One thus expects an AFM spin ordering could take place at above 400 K in $Ak$Th$_2$Mn$_4$P$_4$N$_2$. Future investigations with neutron diffractions will be able to clarify this issue.

The temperature dependence of the specific heat is shown in Figure 3. The specific-heat values at room temperature just exceed $3NR = 324$ J mol$^{-1}$ K$^{-1}$ ($N$ is the number of atoms in a formula unit, and $R$ is the gas constant) expected from the Dulong-Petit law, suggesting other contributions including the electronic part and the magnetic part, in addition to the dominant phonon contribution. No obvious anomaly associated with a magnetic ordering can be observed below 300 K (the bumps near room temperature come from the grease which connects samples with the holder). This result implies that the expected AFM transition could appear above room temperature.

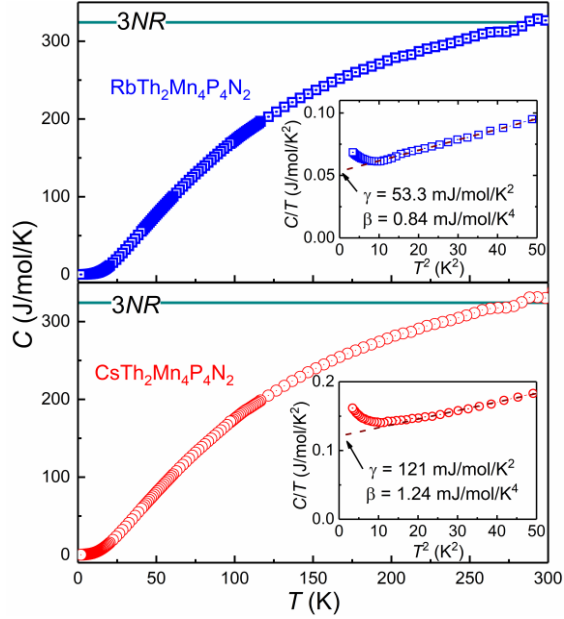

**Figure 3** Temperature-dependent specific heat for RbTh$_2$Mn$_4$P$_4$N$_2$ (a) and CsTh$_2$Mn$_4$P$_4$N$_2$ (b). The Insets show $C/T$ as a function of $T^2$ in the low-temperature region, from which the electronic specific-heat coefficient $\gamma$ is obtained.

As is known, the low-temperature specific heat from lattice obeys Debye's law, namely, $C_\mathrm{lat} = \beta T^3$. Also the specific heat from AFM magnons (albeit being negligibly small at $T \ll T_\mathrm{N}$) is also proportional to $T^3$ [31]. Therefore, the low-temperature specific heat can be fitted with the formula $C = \gamma T + \beta T^3$ (the upturn in $C/T$ below 3 K is probably attributed to a Schottky anomaly). The derived Sommerfeld coefficients are 53.3 mJ mol$^{-1}$ K$^{-2}$ and 121 mJ mol$^{-1}$ K$^{-2}$ for RbTh$_2$Mn$_4$P$_4$N$_2$ and CsTh$_2$Mn$_4$P$_4$N$_2$, respectively, indicating a metallic ground state. Note that the $\gamma$ value of CsTh$_2$Mn$_4$P$_4$N$_2$ (30.3 mJ mol-Mn$^{-1}$ K$^{-2}$) is remarkably larger than those of the related compounds, e.g., 4.2 mJ mol-Mn$^{-1}$ K$^{-2}$ for Ba$_{0.95}$K$_{0.05}$Mn$_2$As$_2$ [28], 8.11 mJ mol$^{-1}$ K$^{-2}$ for ThMnPN [15], and 8.3 mJ mol$^{-1}$ K$^{-2}$ for MnP [33], suggesting enhanced electron correlations. For the small-band-gap AFM semiconductor BaMn$_2$As$_2$, the $\gamma$ value is very close to zero [10], because of the null density of states (DOS) at the Fermi energy. According to $\gamma = \pi^2 N_\mathrm{A} k_\mathrm{B}^2 N_\mathrm{EF}/3$, where $N_\mathrm{A}$ is Avogadro constant and $k_\mathrm{B}$ is the Boltzmann constant, the DOS at $E_\mathrm{F}$ can be derived as $N_\mathrm{EF}$ = 22.58 and 53.42 states/eV for RbTh$_2$Mn$_4$P$_4$N$_2$ and CsTh$_2$Mn$_4$P$_4$N$_2$, respectively. Finally, with the fitted values of $\beta$ and using the formula $\theta_\mathrm{D} = [(12/5) NR\pi^4/\beta]^{1/3}$, the Debye temperature $\theta_\mathrm{D}$'s are estimated to be 311 and 273 K for RbTh$_2$Mn$_4$P$_4$N$_2$ and CsTh$_2$Mn$_4$P$_4$N$_2$, respectively, which are close to that of BaMn$_2$As$_2$ (264 K) [10].

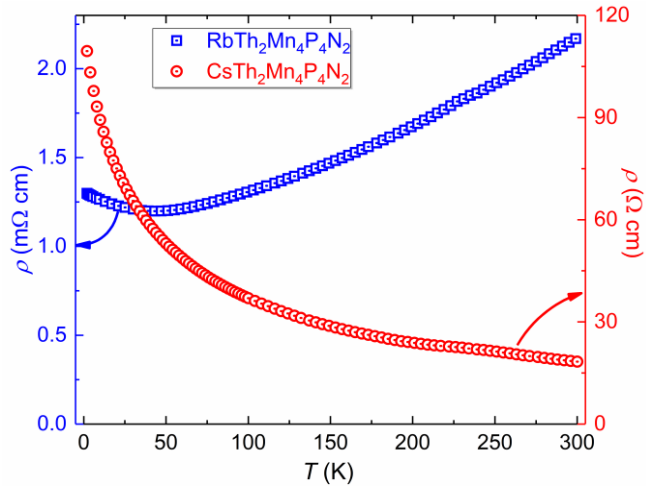

**Figure 4** Temperature dependence of resistivity for the RbTh$_2$Mn$_4$P$_4$N$_2$ (left axis) and CsTh$_2$Mn$_4$P$_4$N$_2$ (right axis) polycrystalline samples.

Figure 4 shows the temperature-dependent electrical resistivity of $Ak$Th$_2$Mn$_4$P$_4$N$_2$ polycrystalline samples. The resistivity of RbTh$_2$Mn$_4$P$_4$N$_2$ at room temperature is 2.2 m$\Omega$ cm, at least six orders of magnitude lower than LaMnPO [4,8,11], and it is even much smaller than those of metallic ThMnPN [14], Ba$_{0.95}$K$_{0.05}$Mn$_2$As$_2$ [28], and La$_{1-x}$Sr$_x$MnAsO [12]. Also, the resistivity decreases with decreasing temperature, which verifies the metallicity. A slight resistivity upturn below 50 K could be due to an electronic localization effect. In the case of $Ak$ = Cs, however, the room-temperature resistivity is much larger (yet it is still one order of magnitude lower than that of LaMnPO). Upon cooling, the resistivity shows a semiconducting-like behavior. Nevertheless, the resistivity at 2 K is not very high (110 Ω cm). Neither Arrhenius relation nor variable-range-hopping conduction can describe the temperature dependence, suggesting that the material could be intrinsically metallic. Indeed, the Fermi level locates at the valence-band tails (see Figs. 5(c) and (f) below), which makes Anderson localization likely. The difference in resistivity between RbTh$_2$Mn$_4$P$_4$N$_2$ and CsTh$_2$Mn$_4$P$_4$N$_2$ is presumably due to the difference in the electron correlations, as revealed by the specific-heat measurements. CsTh$_2$Mn$_4$P$_4$N$_2$ shows pronounced electron correlations, making the itinerant carriers easier to be localized.

**Table 3.** Calculated relative total energies and Mn magnetic moments for RbTh$_2$Mn$_4$P$_4$N$_2$ and CsTh$_2$Mn$_4$P$_4$N$_2$, with different $U$ parameters and under different magnetic configurations (see the text for the definitions of the abbreviations).

| | RbTh$_2$Mn$_4$P$_4$N$_2$ | | | | | |
|---|---|---|---|---|---|---|
| Magnetic Configurations | $U = 0$ | | $U = 1$ eV | | $U = 3$ eV | |
| | $E - E_{NSP}$ (eV/Mn) | $\mu_{Mn}$ ($\mu_B$) | $E - E_{NSP}$ (eV/Mn) | $\mu_{Mn}$ ($\mu_B$) | $E - E_{NSP}$ (eV/Mn) | $\mu_{Mn}$ ($\mu_B$) |
| NSP | 0 | -- | 0 | -- | 0 | -- |
| FM | -0.372 | 2.63 | -0.840 | 3.42 | -2.140 | 4.17 |
| A type | -0.379 | 2.27 | -0.833 | 3.42 | -2.128 | 4.18 |
| C type | Converge to NSP | | -0.019 | 0.16 | -2.484 | 4.01 |
| striped | -0.505 | 3.21 | -0.030 | 0.14 | -2.367 | 4.02 |

| | CsTh$_2$Mn$_4$P$_4$N$_2$ | | | | | |
|---|---|---|---|---|---|---|
| Magnetic Configurations | $U = 0$ | | $U = 1$ eV | | $U = 3$ eV | |
| | $E - E_{NSP}$ (eV/Mn) | $\mu_{Mn}$ ($\mu_B$) | $E - E_{NSP}$ (eV/Mn) | $\mu_{Mn}$ ($\mu_B$) | $E - E_{NSP}$ (eV/Mn) | $\mu_{Mn}$ ($\mu_B$) |
| NSP | 0 | -- | 0 | -- | 0 | -- |
| FM | -0.387 | 2.47 | -0.856 | 3.43 | -2.174 | 4.17 |
| A type | -0.390 | 2.26 | -0.851 | 3.42 | -2.164 | 4.17 |
| C type | Converge to NSP | | -0.014 | 0.16 | -2.506 | 4.02 |
| striped | Converge to NSP | | -0.014 | 0.09 | -2.386 | 4.01 |

To gain insights into the physical origins of the phenomena in RbTh$_2$Mn$_4$P$_4$N$_2$ and CsTh$_2$Mn$_4$P$_4$N$_2$, we performed the first-principles investigations. We first calculated the energy and the magnetic moments of the possible magnetically ordered states including ferromagnetism (FM), interlayer AFM and in-plane FM (A type), in-plane AFM and interlayer FM (C type), and in-plane striped AFM. Non-spin-polarized (NSP) case was also calculated for reference. As shown in Table 3, without onsite Coulomb interaction correction $U$, RbTh$_2$Mn$_4$P$_4$N$_2$ exhibits a striped AFM ground state, while CsTh$_2$Mn$_4$P$_4$N$_2$ has an A-type one. Both compounds cannot hold a stable in-plane AFM state, and tend to have an AFM interlayer coupling (since the A-type AFM has lower energy than FM state). When we switch on the Coulomb interaction correction with $U = 1$ eV, however, intralayer FM dominates. If $U$ is increased to 3 eV, the ground states for both RbTh$_2$Mn$_4$P$_4$N$_2$ and CsTh$_2$Mn$_4$P$_4$N$_2$ become a C-type or Neel type (not calculated). Note that the C-type AFM order has been observed in many other MnP-layer containing compounds (such as ThMnPN [15] and CsMn$_2$P$_2$ [23]), which show similar experimental magnetic behaviors. It is thus reasonable to speculate that such a C-type AFM order also exists in RbTh$_2$Mn$_4$P$_4$N$_2$ and CsTh$_2$Mn$_4$P$_4$N$_2$. If this is the case, the onsite Coulomb interactions should be significant, given that the intralayer AFM order is favored at a large value of $U$ in RbTh$_2$Mn$_4$P$_4$N$_2$ and CsTh$_2$Mn$_4$P$_4$N$_2$. Expectedly, the calculated magnetic moment increases with $U$, and the Mn magnetic moment at $U = 3$ eV turns out to be 4.0 $\mu_B$, which is very close to the experimental results in BaMn$_2$As$_2$ (3.9 $\mu_B$ at 10 K) [10] and ThMnPN (3.6 $\mu_B$ at 4 K) [15].

Here we perform fittings to the energies of the $U = 3$ eV case, with a simple Heisenberg model containing the interactions of intralayer nearest-neighbor ($J_1$), intralayer next-nearest-neighbor ($J_2$), and interlayer nearest-neighbor ($J_3$) [31]. The energies of FM, A-type AFM, C-type AFM, and striped AFM can be written as

$E_{FM} = E_0 + (2J_1 + 2J_2 + J_3)S^2$,
$E_A = E_0 + (2J_1 + 2J_2 - J_3)S^2$,
$E_C = E_0 + (-2J_1 + 2J_2 + J_3)S^2$,
$E_S = E_0 + (-2J_2 + J_3)S^2$,

where $E_0$ is an energy offset, $S$ is the (classical) local spin, and all the energies in the values of meV per Mn moment. The solutions are, $J_1 = 1/4 (E_{FM} - E_C)/S^2$, $J_2 = 1/4 (E_C - E_S + 2J_1S^2)/S^2$, and $J_3 = 1/2 (E_{FM} - E_A)/S^2$. Assuming $S = 2.0$ (corresponding to the calculated moment of 4.0 $\mu_B$/Mn), the exchange interactions are $J_1 = 21.5$ meV, $J_2 = 3.43$ meV, $J_3 = -1.45$ meV for RbTh$_2$Mn$_4$P$_4$N$_2$, and $J_1 = 20.7$ meV, $J_2 = 2.88$ meV, $J_3 = -1.28$ meV for CsTh$_2$Mn$_4$P$_4$N$_2$. The large positive $J_1$ and $J_1/J_2$ values indicate dominant AFM orders with a high Neel temperature, while the small negative $J_3$ values suggest weak ferromagnetic couplings between adjacent MnP layers. It is notable that all $J$'s for RbTh$_2$Mn$_4$P$_4$N$_2$ are slightly larger than those for CsTh$_2$Mn$_4$P$_4$N$_2$. Such a difference is consistent with the shorter Mn–P bond distances in RbTh$_2$Mn$_4$P$_4$N$_2$, indicating stronger hybridization between Mn-3$d$ and P-3$p$ states in RbTh$_2$Mn$_4$P$_4$N$_2$.

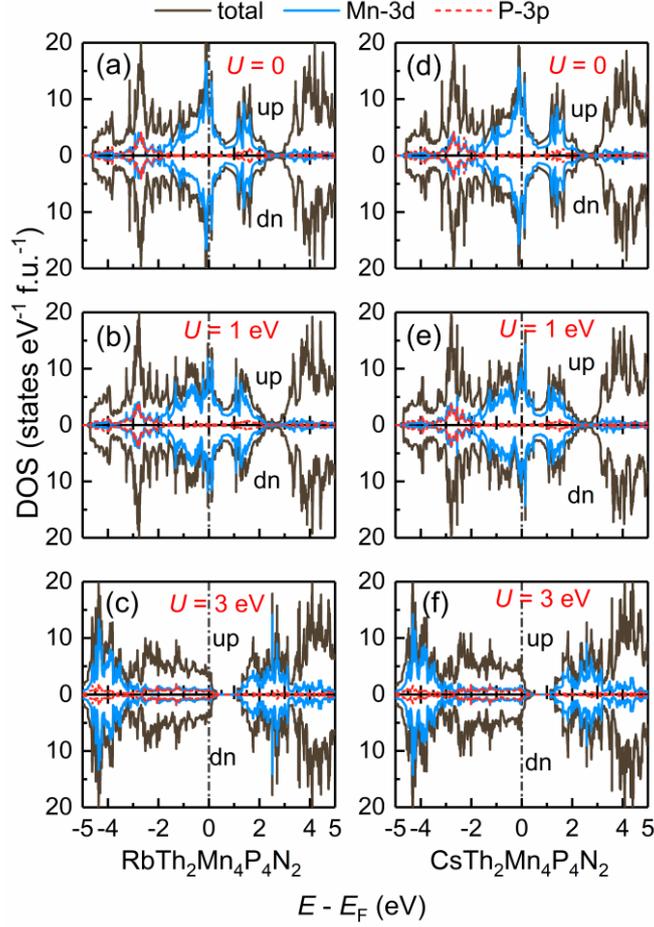

**Figure 5** Calculated density of states for RbTh$_2$Mn$_4$P$_4$N$_2$ and CsTh$_2$Mn$_4$P$_4$N$_2$ with different setups: (a, d) $U = 0$, non-spin-polarized; (b, e) $U = 1$ eV, C-type AFM; (c, f) $U = 3$ eV, C-type AFM.

The hybridization properties between Mn-3$d$ and P-3$p$ states can be further comprehended with the calculations of the projected density of states (DOS). As shown in Fig. 5, RbTh$_2$Mn$_4$P$_4$N$_2$ and CsTh$_2$Mn$_4$P$_4$N$_2$ exhibit quite similar electronic structures. In the case of $U = 0$, both NSP RbTh$_2$Mn$_4$P$_4$N$_2$ and CsTh$_2$Mn$_4$P$_4$N$_2$ have metallic band structures, and the DOS values at Fermi level ($N^0_{EF}$) are 18.79 eV$^{-1}$ f.u.$^{-1}$ and 18.81 eV$^{-1}$ f.u.$^{-1}$ for NSP RbTh$_2$Mn$_4$P$_4$N$_2$ and CsTh$_2$Mn$_4$P$_4$N$_2$, respectively. At $U = 1$ eV, the DOS structures of the Neel magnetic order state are very similar to those of the NSP state. In both cases of $U = 0$ and $U = 1$ eV, the DOS's at Fermi level are dominantly contributed by the Mn-3$d$ states, and the $d$-$p$ hybridization is very weak.

However, when $U$ is increased to 3 eV, the electronic structure drastically changes as the Neel-type ground state develops. As can be seen in Fig. 5 (c) and (f), the Mn-3$d$ contributions are significantly reduced while P-3$p$ contributions are increased. Both $d$ and $p$ partial DOS's have similar energy dependencies. Such results demonstrate strong $d$-$p$ hybridization for the $U = 3$ eV case. The strong hybridization picture has also been proposed in other Mn$Pn$-layer-containing compounds [4,8,14,34].

There emerges an energy gap of $\sim 0.7$ eV, slightly above Fermi level for $U = 3$ eV. Integration of DOS shows that the gap lies exactly at the filling level of the nominal 3$d^5$ configuration, indicating a close relationship to the Mott physics [35]. Note that the $d$-$p$ hybridization is significant for the lower Hubbard bands, and the Fermi level locates at the band tails. This means that the doped holes partially reside the P site, consistent with the crystal-structure trend above. Moreover, the $N^0_{EF}$ values for RbTh$_2$Mn$_4$P$_4$N$_2$ and CsTh$_2$Mn$_4$P$_4$N$_2$ in the case of $U = 3$ eV decrease to 6.46 eV$^{-1}$ f.u.$^{-1}$ and 7.10 eV$^{-1}$ f.u.$^{-1}$, which are much lower than the values of 22.58 eV$^{-1}$ f.u.$^{-1}$ and 53.42 eV$^{-1}$ f.u.$^{-1}$ derived from the specific heat measurements. This discrepancy suggests remarkable renormalization effects from electron-electron and electron-phonon interactions, according to the formula $N_{EF} = N^0_{EF} (1 + \lambda_{ep} + \lambda_{ee})$, where $\lambda_{ep}$ and $\lambda_{ee}$ denote the electron-electron and electron-phonon coupling. Since the electron-phonon coupling constant is normally smaller than 2 [36], the large $N_{EF}/N^0_{EF}$ values (3.5 and 7.5 for RbTh$_2$Mn$_4$P$_4$N$_2$ and

CsTh$_2$Mn$_4$P$_4$N$_2$, respectively) point to significant electron correlations especially for CsTh$_2$Mn$_4$P$_4$N$_2$. Note that there are strong electron correlation effects in Ba$_{0.95}$K$_{0.05}$Mn$_2$As$_2$ although the Sommerfeld constant is much lower [28].

**Discussion**

Now let us briefly discuss the underlying physics of the structural and physical properties of $Ak$Th$_2$Mn$_4$P$_4$N$_2$. First of all, the inter-block-layer charge transfer makes the ThMnAsN block layers shrink, which finally leads to the unusual reduction of the $a$ axes of $Ak$Th$_2$Mn$_4$P$_4$N$_2$. The charge transfer is also reflected by the positional adjustment of P atoms along the $c$ direction. Secondly, the '$c/a$' parameter for the ThMnAsN block layers remains to be small, indicating that the chemical pressure from the [Th$_2$N$_2$] layers is preserved. It was found in our previous studies [6,15,37,38] that materials containing [Th$_2$N$_2$] layers exhibit different properties from their analogous compounds, primarily due to their exclusive chemical pressure effect. For example, ThFeAsN is the unique 1111-type superconductor by itself (without extrinsic doping) [37]. The chemical pressure may enhance the Mn3$d$-P3$p$ hybridizations, and the doped holes have more P-3$p$ properties.

Our first-principles calculations indicate that the Fermi level locates at the top of the lower Hubbard bands, which suggests metallic conductions from the doped holes. While RbTh$_2$Mn$_4$P$_4$N$_2$ indeed shows a metallicity from the electrical resistivity data, CsTh$_2$Mn$_4$P$_4$N$_2$ apparently exhibits a semiconducting-like behavior. This is tentatively explained in terms of enhanced electron correlations in CsTh$_2$Mn$_4$P$_4$N$_2$ that makes the doped holes easily localized. Nevertheless, the semiconducting-like behavior in CsTh$_2$Mn$_4$P$_4$N$_2$ may also be caused by an extrinsic origin. One possibility is that the sample deviates from its stoichiometry (say, with N deficiency) which makes the conduction holes compensated. An additional possibility is that, for some reasons, the sample has thick semiconducting grain boundaries.

Although no magnetic ordering was detected up to 400 K by the magnetic susceptibility measurement, one expects an AFM transition at a higher temperature because the effective moments derived from the CW fit are too small to account for the Mn-3$d$ magnetism. The effective localized moments of $Ak$Th$_2$Mn$_4$P$_4$N$_2$ are close to that of ThMnPN [15]. Note that the Neel temperatures of the related materials are $T_N > 300$ K for ThMnPN [15] and, $T_N > 750$ K for BaMn$_2$P$_2$ [7]. Our theoretical calculations also suggest AFM interactions with a high Neel temperature. In a word, $Ak$Th$_2$Mn$_4$P$_4$N$_2$ appears to involve dominant localized electrons with a local moment of 4.0 $\mu_B$/Mn, which couple with the itinerant electrons, which gives rise to a novel AFM correlated metallic state.

## Conclusions

In summary, we have designed, synthesized and characterized two new compounds $Ak$Th$_2$Mn$_4$P$_4$N$_2$ with separate [Mn$_2$P$_2$] bilayers. Lattice match and charge transfer play the crucial role in stabilizing the intergrowth structure. It was found that the interlayer charge transfer is responsible for the reduction of $a$ axis, while the internal chemical pressure remains. The physical property measurements in combination with the first-principles calculations indicate that the new materials simultaneously host antiferromagnetically coupled local moments, itinerant holes with P-3$p$ character, and strong electron correlations. Among them, the robust antiferromagnetism seems to be the main obstacle for finding superconductivity in Mn-based layered compounds.

## Experimental

The target polycrystalline samples of $Ak$Th$_2$Mn$_4$P$_4$N$_2$ ($Ak$ = Rb, Cs) were synthesized by solid-state reaction with powder of Th$_3$N$_4$, Th, MnP and $Ak$Mn$_2$P$_2$ as starting materials. The preparation of the precursors (Th, Th$_3$N$_4$) is similar to the ThFeAsN parent compound [37]. MnP was prepared by heating the mixture of Manganese (99.9% Alfa Aesar) and Phosphorus (99.99% Alfa Aesar) powders in an evacuated quartz ampule at 973 K for 48 h. $Ak$Mn$_2$P$_2$ was additionally prepared by reacting Rubidium ingot (99.75% Alfa Aesar) / Cesium ingot (99.5% Alfa Aesar) and MnP at 973 K for 24 h (The XRD can be seen in Support Information Figure. S2). With these intermediate precursors, $Ak$Th$_2$Mn$_4$P$_4$N$_2$ samples were finally synthesized by solid-state reactions of the stoichiometric mixtures. The chemical reactions take place in a small alumina container which is sealed in a Ta tube, and the Ta tube was jacketed with an evacuated quartz ampule. This sample-loaded ampule was sintered at 1223 K for 40 h, after which it was allowed to cool down by switching off the furnace. Then, the solid-state reaction was repeated, with a sample homogenization by grinding to improve the quality of the sample. The final product was black in color and stable in air. All the operations above were carried out in an argon-filled glovebox with the water and oxygen content below 1 ppm.

Powder x-ray diffraction (XRD) was carried out at room temperature on a PANalytical X-ray diffractometer (Model EMPYREAN) with a monochromatic Cu K$\alpha_1$ radiation. Crystal structure data were obtained by Rietveld refinement using the step-scan XRD data with $15° < \theta < 150°$ for this sample. Magnetic measurements were performed on a Quantum Design Magnetic Property Measurement System (MPMS3), and the sample was cut into a rectangular shape with the intension of diminish the demagnetization factor. The temperature-dependent resistivity was measured using a standard four-terminal method on a Physical Property Measurement System (PPMS) equipped with a Keithley 2400 digital sourcemeter and a Keithley 2182 nanovoltmeter. The current in the measurements was reversed for each data point to remove the thermoelectric effects. Measurement of the specific heat was performed on a Quantum Design Physical Property Measurement System (PPMS-9) using a thermal relaxation method. Our density functional theory (DFT) calculations were performed with projected augmented wave (PAW) basis [39], as implemented in the VASP package [40]. The exchange-correlation energy was treated using the Perdew, Burke, and Enzerhoff-type generalized-gradient-approximation (GGA) functionals [41], with rotationally invariant onsite Coulomb interaction corrections ($+U$) [42]. For all calculations, we adopted the experimental structures. The plane wave energy cutoff was set 600 eV. The k-mesh was set 15×15×2 gamma-centered for self-consistent and 30×30×4 for density of states calculations.

## Supporting Information

The supporting information (the energy-dispersive x-ray spectra data of $Ak$Th$_2$Mn$_4$P$_4$N$_2$ ($Ak$ = Rb, Cs) and X-ray diffraction of intermediate precursors) for this article is available on the WWW under https://doi.org/10.1002/cjoc.2021xxxxx.

## Acknowledgement

This work was supported by the National Key Research and Development Program of China (2017YFA0303002), National Natural Science Foundation of China (12050003), and the Key R&D Program of Zhejiang Province, China (2021C01002).

## References

[1] Kamihara, Y.; Watanabe, T.; Hirano, M.; and Hosono, H. Iron-Based Layered Superconductor La[O$_{1-x}$F$_x$]FeAs ($x$ = 0.05−0.12) with $T_c$ = 26 K. *J. Am. Chem. Soc.* **2008**, *130*, 3296–3297.

[2] Chen, X. H.; Wu, T.; Wu, G.; Liu, R. H.; Chen, H. & Fang, D. F. Superconductivity at 43 K in SmFeAsO$_{1-x}$F$_x$. *Nature* **2008**, *453*, 761–762.

[3] Xu, G.; Ming, W.; Yao, Y.; Dai, X.; Zhang, S.-C. and Fang, Z. Doping-dependent phase diagram of LaO$M$As ($M$ = V – Cu) and electron-type superconductivity near ferromagnetic instability. *EPL* **2008**, *82*, 67002.

[4] Yanagi, H.; Watanabe, T.; Kodama, K.; Likubo, S.; Shamoto, S.; Kamiya, T.; Hirano, M.; and Hosono, H. Antiferromagnetic bipolar semiconductor LaMnPO with ZrCuSiAs-type structure. *J. Appl. Phys.* **2009**, *105*, 093916.

[5] Park, S.-W.; Mizoguchi, H.; Kodama, K.; Shamoto, S.; Otomo, T.; Matsuishi, S.; Kamiya, T.; and Hosono, H. Magnetic Structure and Electromagnetic Properties of $Ln$CrAsO with a ZrCuSiAs-type Structure ($Ln$ = La, Ce, Pr, and Nd). *Inorg. Chem.* **2013**, *52*, 13363–13368.

[6] Wang, Z.-C.; Shao, Y.-T.; Wang, C.; Wang Z.; Xu, Z.-A. and Cao, G.-H. Enhanced superconductivity in ThNiAsN. *EPL* **2017**, *118*, 57004.

[7] Brock, L. S.; Greedan, E. J.; Kauzlarich, M. S. Resistivity and Magnetism of $A$Mn$_2$P$_2$ ($A$ = Sr, Ba): The Effect of Structure Type on Physical Properties. *J. Solid State Chem.* **1994**, *113*, 303-311.

[8] Simonson, W. J.; Yin, Z. P.; Pezzoli, M.; Guo, J.; Liu, J.; Post, K.; Efimenko, A.; Hollmann, N.; Hu, Z.; Lin, H.-J.; Chen, C.-T.; Marques, C.; Leyva, V.; Smith, G.; Lynn, W. J.; Sun, L. L.; Kotliar, G.; Basov, N. D.; Tjeng, H. L.; and Aronson, C. M. From antiferromagnetic insulator to correlated metal in pressurized and doped LaMnPO. *PNAS* **2012** *109*, 1815-1819.

[9] Guo, J.; Simonson, W. J.; Sun, L.; Wu, Q.; Gao, P.; Zhang, C.; Gu, D.; Kotliar, G.; Aronson, M. & Zhao, Z.-X. Observation of antiferromagnetic order collapse in the pressurized insulator LaMnPO. *Sci. Rep.* **2013**, *3*, 2555.

[10] Singh, Y.; Green, A. M.; Huang, Q.; Kreyssig, A.; McQueeney, J. R.; Johnston, C. D.; and Goldman, I. A. Magnetic order in BaMn$_2$As$_2$ from neutron diffraction measurements. *Phys. Rev. B* **2009**, *80*, 100403(R).

[11] Simonson, W. J.; Post, K.; Marques, C.; Smith, G.; Khatib, O.; Basov, N. D.; and Aronson, C. M. Gap states in insulating LaMnPO$_{1-x}$F$_x$ ($x$ = 0 - 0.3). *Phys. Rev. B* **2011**, *84*, 165129.

[12] Sun, Y.-L.; Bao, J.-K.; Luo, Y.-K.; Feng, C.-M.; Xu, Z.-A. and Cao, G.-H. Insulator-to-metal transition and large thermoelectric effect in La$_{1-x}$Sr$_x$MnAsO. *EPL* **2012**, *98*, 17009.

[13] Wildman, J. E.; Emery, N.; and Mclaughlin, C. A. Electronic and magnetic properties of Nd$_{1-x}$Sr$_x$MnAsO oxyarsenides. *Phys. Rev. B* **2014**, *90*, 224413.

[14] Bao, J.-K.; Jiang, H.; Sun, Y.-L.; Jiao, W.-H.; Shen, C.-Y.; Guo, H.-J.; Chen, Y.; Feng, C.-M.; Yuan, H.-Q.; Xu, Z.-A.; Cao, G.-H.; Sasaki, R.; Tanaka, T.; Matsubayashi, K.; and Uwatoko, Y. Weakly ferromagnetic metallic state in heavily doped Ba$_{1-x}$K$_x$Mn$_2$As$_2$. *Phys. Rev. B.* **2012**, *85*, 144523.

[15] Zhang, F.-X.; Li, B.-Z.; Ren, Q.-Y.; Mao, H.-C.; Xia, Y.; Hu, B.; Liu, Z.; Wang, Z.-C.; Shao, Y.-T.; Feng, Z.; Tan, S.-G.; Sun, Y.-P.; Ren, Z.; Jing, Q.; Liu, B.; Luo, H.-C.; Ma, J.; Mei, Y.-X.; Wang, C.; and Cao, G.-H. ThMn$Pn$N ($Pn$ = P, As): Synthesis, Structure, and Chemical Pressure Effects. *Inorg. Chem.* **2020**, *59*, 2937−2944.

[16] Cheng, J.-G.; Matsubayashi, K.; Wu, W.; Sun, J.P.; Lin, F.K.; Luo, J.L.; and Uwatoko, Y. Pressure Induced Superconductivity on the border of Magnetic Order in MnP. *Phys. Rev. Lett.* **2015**, *114*, 117001.

[17] Wang, Z.-C.; He, C.-Y.; Wu, S.-Q.; Tang, Z.-T.; Liu, Y.; Abduweli, A.; Feng, C.-M.; and Cao, G.-H. Superconductivity in KCa$_2$Fe$_4$As$_4$F$_2$ with Separate Double Fe$_2$As$_2$ Layers. *J. Am. Chem. Soc.* **2016**, *138*, 7856–7859.

[18] Wang, Z.-C.; He, C.Y.; Tang, Z.-T.; Wu, S.-Q.; and Cao, G.-H. Crystal structure and superconductivity at about 30 K in $A$Ca$_2$Fe$_4$As$_4$F$_2$ ($A$ = Rb, Cs). *Sci. China Mater.* **2017**, *60*, 83–89.

[19] Wang, Z.-C.; He, C.-Y.; Wu, S.-Q.; Tang, Z.-T.; Liu, Y.; and Cao, G.-H. Synthesis, Crystal Structure and Superconductivity in Rb$Ln_2$Fe$_4$As$_4$O$_2$ ($Ln$ = Sm, Tb, Dy and Ho). *Chem. Mater.* **2017**, *29*, 1805–1812.

[20] Wu, S.-Q.; Wang, Z.-C.; He, C.-Y.; Tang, Z.-T.; Liu, Y.; and Cao, G.-H. Superconductivity at 33 – 37 K in $ALn_2$Fe$_4$As$_4$O$_2$ ($A$

= K and Cs, *Ln* = lanthanides). *Phys. Rev. Materials* **2017**, *1*, 044804.

[21] Shao, Y.-T.; Wang, Z.-C.; Li, B.-Z.; Wu, S.-Q.; Wu, J.-F.; Ren, Z.; Qiu, S.-W.; Rao, C.; Wang, C. & Cao, G.-H. BaTh$_2$Fe$_4$As$_4$(N$_{0.7}$O$_{0.3}$)$_2$: An iron-based superconductor stabilized by inter-block-layer charge transfer. *Sci. China Mater.* **2019**, *62*, 1357–1362.

[22] Wang, Z.-C.; Wu, S.-Q.; Ji, L.-W. and Cao, G.-H. Block-Layer Model for Intergrowth Structure. *Nano Res.* **2021**, to be published.

[23] Hummel, F.; Johrendt, D. *Ph.D Dissertation*, der Fakultät für Chemie und Pharmazie, der Ludwig-Maximilians-Universität München, **2005**.

[24] Izumi, F. and Momma, K.; Three-Dimensional Visualization in Powder Diffraction. *Solid State Phenom.* **2007**, *130*, 15-20.

[25] Rotter, M.; Pangerl, M.; Tegel, M.; Johrendt, D. Superconductivity and Crystal Structures of (Ba$_{1-x}$K$_x$)Fe$_2$As$_2$ ($x$ = 0–1), *Angew. Chem. Int. Ed.* **2008**, *47*, 7949–7952.

[26] Nientiedt, T. A.; Jeitschko, W.; Pollmeier, G. P. and Brylak, M. Quaternary Equiatomic Manganese Pnictide Oxides *A*MnPO (*A* = La-Nd, Sm, Gd-Dy), *A*MnAsO (*A* = Y, La-Nd, Sm, Gd-Dy, U), and *A*MnSbO (*A* = La-Nd, Sm, Gd) with ZrCuSiAs Type Structure. *Z. Naturforsch.* **1997**, *52*, 560-564.

[27] Na, S-H.; Wu, W. and Luo, J.-L. Anisotropy Properties of Mn$_2$P Single Crystals with Antiferromagnetic Transition. *Chin. Phys. Lett.* **2020**, *37*, 087301.

[28] Pandey, A.; Dhaka, S. R.; Lamsal, J.; Lee, Y.; Anand, K. V.; Kreyssig, A.; Heitmann, W. T.; McQueeney, J. R.; Goldman, I. A.; Harmon, N. B.; Kaminski, A.; and Johnston, C. D. Ba$_{1-x}$K$_x$Mn$_2$As$_2$: An Antiferromagnetic Local-Moment Metal. *Phys. Rev. Lett.* **2012**, *108*, 087005.

[29] Pandey, A.; Ueland, G. B.; Yeninas, S.; Kreyssig, A.; Sapkota, A.; Yang, Z.; Helton, S. J.; Lynn, W. J.; McQueeney, J. R.; urukawa, Y. F.; Goldman, I. A.; and Johnston, C. D. Coexistence of Half-Metallic Itinerant Ferromagnetism with Local-Moment Antiferromagnetism in Ba$_{0.60}$K$_{0.40}$Mn$_2$As$_2$. *Phys. Rev. Lett.* **2018**, *111*, 047001.

[30] Glasbrenner, K. J.; and Mazin, I. I. First-principles evidence of Mn moment canting in hole-doped Ba$_{1-2x}$K$_{2x}$Mn$_2$As$_2$. *Phys. Rev. B* **2014**, *89*, 060403(R).

[31] Johnston, C. D.; McQueeney, J. R.; Lake, B.; Honecker, A.; Zhitomirsky, E. M.; Nath, R.; Furukawa, Y.; Antropov, P. V.; and Singh, Y. Magnetic exchange interactions in BaMn$_2$As$_2$: A case study of the $J_1$-$J_2$-$J_c$ Heisenberg model, *Phys. Rev. B* **2011**, *84*, 094445.

[32] Marcinkova, A.; Hansen, T. C.; Curfs, C.; Margadonna, S.; Bos, J.-W. G. Nd-induced Mn spin-reorientation transition in NdMnAsO. *Phys. Rev. B* **2010**, *82*, 174438.

[33] Zheng, P.; Xu, Y. J.; Wu, W.; Xu, G.; Lv, J. L.; Lin, F. K.; Wang, P.; Yang, Y.-F. & Luo, J. L. Orbital-dependent charge dynamics in MnP revealed by optical study. *Sci. Rep.* **2017**, *7*, 14178.

[34] An, J.-M.; Sefat, S. A.; Singh, J. D.; and Du, M.-H. Electronic structure and magnetism in BaMn$_2$As$_2$ and BaMn$_2$Sb$_2$. *Phys. Rev. B* **2009**, *79*, 075120.

[35] Mott, F. N. Metal-insulator transitions; *Contemp. Phys.*, **1973**, *14*, 401-413.

[36] Abduweli, A.; Sun, Y.-L.; Cheng, E.-J.; Liu, Y.-B.; Wu, S.-Q.; Jiang, H.; Ren, Z.; Li, S.-Y.; and Cao, G.-H. V$_2$Te$_2$O: A Two-Dimensional van der Waals Correlated Metal. *Inorg. Chem.* **2018**, *57*, 14617−14623.

[37] Wang, C.; Wang, Z.-C.; Mei, Y.-X.; Li, Y.-K.; Li, L.; Tang, Z.-T.; Liu, Y.; Zhang, P.; Zhai, H.-F.; Xu, Z.-A.; and Cao, G.-H. A New ZrCuSiAs-Type Superconductor: ThFeAsN. *J. Am. Chem. Soc.* **2016**, *138*, 2170–2173.

[38] Li, B.-Z.; Wang, Z.-C.; Wang, J.-L.; Zhang, F.-X.; Wang, D.-Z.; Zhang, F.-Y.; Sun, Y.-P.; Jing, Q.; Zhang, H.-F.; Tan, S.-G. Li, Y.-K.; Feng, C.-M.; Mei, Y.-X.; Wang, C. and Cao, G.-H. Peculiar phase diagram with isolated superconducting regions in ThFeAsN$_{1-x}$O$_x$. *J. Phys.: Condens. Matter* **2018**, *30*, 255602.

[39] Blöchl, E. P. Projector augmented-wave method. *Phys. Rev. B.* **1994**, *50*, 17953.

[40] Kresse, G. and Furthmüller, J. Efficient iterative schemes for *ab* initio total-energy calculations using a plane-wave basis set. *Phys. Rev. B.* **1996**, *54*, 11169.

[41] Perdew, P. J.; Burke, K.; and Ernzerhof, M. Generalized Gradient Approximation Made Simple. *Phys. Rev. Lett.*, **1996**, *77*, 3865.

[42] Dudarev, L. S.; Botton, A. G.; Savrasov, Y. S.; Humphreys, J. C. and Sutton, P. A. Electron-energy-loss spectra and the structural stability of nickel oxide: An LSDA+*U* study. *Phys. Rev. B*, **1998**, *57*, 1505.